\documentclass[superscriptaddress,nofootinbib,showpacs,preprint,preprintnumbers,amsmath,amssymb]{revtex4-1}

\usepackage{amsmath}
\usepackage{epsfig}
\usepackage{graphicx,subfigure}
\usepackage{xcolor}
\usepackage{latexsym}
\usepackage{amssymb}
\usepackage{ulem}

\def\nn{\nonumber}

\def\GeV{\,{\rm GeV}}
\def\TeV{\,{\rm TeV}}
\def\fb{\,{\rm fb}}
\def\invfb{\,{\rm fb}^{-1}}
\def\Br{{\cal B}}
\def\lamct{\lambda_{ct}}
\def\lamtt{\lambda_{tt}}

\newcommand{\tg}[1]{(#1)}

\begin{document}

\baselineskip 3.0ex
\vspace*{18pt}

\begin{flushright}
\end{flushright}

\title{Exploring top quark FCNC within 2HDM type III \\
in association with flavor physics}

\author{C.~S.~Kim}
\email{cskim@yonsei.ac.kr }
\affiliation{Department of Physics and IPAP, Yonsei University, Seoul 120-749, Korea}
\author{Yeo Woong Yoon}
\email{ywyoon@kias.re.kr}
\affiliation{School of Physics, KonKuk University, Seoul 143-701, Korea}
\affiliation{School of Physics, Korea Institute for Advanced Study, Seoul 130-722, Korea}
\author{Xing-Bo Yuan }
\email{xbyuan@yonsei.ac.kr}
\affiliation{Department of Physics and IPAP, Yonsei University, Seoul 120-749, Korea}

\begin{abstract}
\vspace{0.3cm}
\baselineskip 3.0ex  \noindent The top quark flavor
changing neutral current (FCNC) process is an excellent probe to
search for new physics in top sector since the Standard Model
expectation is extremely suppressed. We explore Higgs-mediated top
quark FCNC, focusing on $H$-$t$-$c$ Yukawa coupling $\lambda_{ct}$ within
the general two Higgs doublet model. After electroweak symmetry
breaking the top quark FCNC couplings are included in the charged
Higgs Yukawa sector so that they contribute to various processes
in flavor physics.
 To probe $\lambda_{ct}$, we study anomalous single top production and the same sign top pair production at the LHC in association with flavor physics from the tree-level processes  $B\to D^{(*)}\tau\nu$, $B\to \tau \nu$ as well as from the loop-level processes  $B_d \to X_s \gamma$, $B_{d,s}-{\overline B}_{d,s}$ mixing.
 We perform combined analysis of all the constraints regarding the fine-tuning argument
 to fit the data and discuss future prospect. The recently updated measurements on $B\to D^{(*)}\tau\nu$
 still prefer large $\lambda_{ct}$, but we show that the current bound on
 the same sign top pair production at the LHC gives the most significant upper bound on  $\lambda_{ct}$
 to be less than $10\sim30$ depending on neutral heavy Higgs masses. We also find that for the given upper
 bound on $\lambda_{ct}$, $B\to D^{(*)}\tau\nu$ put significant lower bound on $H$-$\tau$-$\tau$ Yukawa coupling,
 and the bound is proportional to the charged Higgs mass.

\end{abstract}

\preprint{KIAS-P15051}

\maketitle

\section{Introduction}

The top quark, the  heaviest particle in the Standard Model (SM),
plays an important role as an input for the electroweak (EW)
precision measurements~\cite{Baak:2014ora}. Because its mass is
much heavier than other known particles, the top quark is considered to
be the most viable candidate which has a close connection to new
physics (NP) that controls the EW symmetry breaking mechanism. Meanwhile, the
discovery of the SM-like Higgs boson at the
LHC~\cite{Aad:2012tfa,Chatrchyan:2012xdj} and the precision
measurement of its
property~\cite{Khachatryan:2014jba,ATLAS:2015HIG1} shed much light
on the physics in EW sector, boosting the relevant studies.
Especially, NP scenarios with extended Higgs sector have received
great interest due to its rich phenomenology and attempt to
complement the SM~\cite{Martin:1997ns,Branco:11}.

 One of the simplest scenarios with extended Higgs sector is to introduce a
new Higgs doublet. Because the two Higgs doublets can couple to
both up-type and down-type quarks, after rotating into their mass
eigenstates, the tree-level flavor changing neutral current (FCNC)
inevitably arises. In the SM, the tree-level FCNC is forbidden by
the GIM mechanism~\cite{Glashow:1970gm}. The FCNC process only
takes place through the loop diagrams with charged current and
rough estimation of the loop correction at the amplitude level is
\begin{equation}
V_{\rm CKM}^\prime  V_{\rm CKM}^* \,\frac{\alpha_e}{4\pi} \Big(\frac{m_q}{m_W}\Big)^2,
\end{equation}
 where $V_{\rm CKM}^{(\prime)}$ are CKM matrices, $m_q$ is the mass of quark inside the loop. Thus, the loop-induced down-type quark FCNC processes such as $b\to s\gamma$, which is involved with top quark loop, has enhancement factor $(m_t/m_W)^2$ and their rates mostly fall within current experimental reach of $B$ physics and Kaon physics. Therefore, the down-type quark FCNC is severely constrained and dangerous to many NP scenarios.
 On the other hand, the up-type quark FCNC processes, for example top quark FCNC process $t\to c\gamma$, are involved with $b$-quark loop and extremely suppressed by $(m_b/m_W)^2$. The estimation of  ${\Br}(t\to c\gamma)$ is ${\cal O}(10^{-12})$~\cite{Eilam:1990zc} within the SM, far too much behind the current experimental reach.

 In order to avoid tree-level FCNC, one usually introduces a discrete $Z_2$ symmetry
 to make each up-type or down-type quark couple to only one Higgs doublet.
 In the Minimal Supersymmetric Standard Model (MSSM) the supersymmetry itself plays the role.
 Without such a $Z_2$ symmetry, the general 2HDM which is called ``2HDM type III'' follows
 a specific scheme to circumvent severe down-type quark FCNC constraints
 such as the natural flavor conservation~\cite{Glashow:76},
 the minimal flavor violation~\cite{Chivukula:87,Buras:20,D'Ambrosio:02,Buras:10,Isidori:12,Cervero:12}
 and Cheng-Sher ansatz~\cite{Cheng:87}. In this work we adopt the last one, in which the Yukawa coupling $\xi_{ij}$ 
 connecting quarks with flavor indices $i,j$ to one of the neutral Higgses
 is described as
 \begin{equation}
 \label{eq:ChengSher}
 \xi_{ij} = \lambda_{ij} \frac{\sqrt{2 m_i m_j}}{v}\,,
 \end{equation}
 where $v$ is the SM vacuum expectation value (vev), $v=246\GeV$, $\lambda_{ij}$ is considered to be ${\cal O}(1)$.
 With this ansatz,  down-type quark FCNC is severely suppressed due to the small masses of $u,d,s$ quarks, being safe against the experimental constraints.
 However, top quark FCNC process can be potentially large and should be explored
 in collider physics as well as in flavor physics.

  In the 2HDM type III, after EW symmetry breaking the top quark FCNC Yukawa couplings $\lambda_{qt}$ ($q=u,\,c$)
  also come into play in charged Higgs Yukawa couplings.
  Therefore, the phenomenology of top quark FCNC process with neutral Higgs exchange is naturally in connection
  with flavor physics process with charged Higgs exchanged due to the common Yukawa couplings $\lambda_{qt}$.
  Studies on the top quark FCNC in collider physics especially through anomalous top quark decays were performed
  in Refs.~\cite{Hou:1991un,CorderoCid:2004vi,Larios:2006pb,Ferreira:2008cj,Aranda:2009cd}.
  There have been studies on the issue that large top quark FCNC coupling $\lambda_{ct}$
  is needed~\cite{Crivellin:12,Chen:13} to explain the measurements of $\Br (B_d \to D^{(*)} \tau \nu)$
  at BaBar~\cite{Lees:2012xj}, which were quite larger than the SM expectations.
  The authors of Ref.~\cite{Gupta:2009wn} study the collider signature with constraints
  from $b\to s\gamma$ concerning the perturbativity of Yukawa couplings within the 2HDM and the MSSM.
  For more comprehensive study on 2HDM type III contribution to both collider and flavor physics,
  we refer to Ref.~\cite{Atwood:96}.
  The model independent approach using low energy effective operators was done in Ref.~\cite{Fox:2007in}.

  In this work we focus on $H$-$t$-$c$ FCNC coupling $\lambda_{ct}$ within 2HDM type III by adopting Cheng-Sher ansatz. We perform detailed study on several experimental observables that can give bound on $\lambda_{ct}$ from collider physics and flavor physics with the most up-to-date experimental data. The issue on $\Br (B_d \to D^{(*)} \tau \nu)$ is revisited with new data from Belle and LHCb. Especially it will be shown that the search for the same sign top pair production at the LHC plays crucial role to constrain $\lambda_{ct}$. Since the current precision measurements of the SM Higgs properties are very well consistent with the SM expectations~\cite{Khachatryan:2014jba,ATLAS:2015HIG1}, we assume the alignment limit for the Higgs potential of 2HDM type III, in which the SM Higgs sector is well decoupled from the NP sector.

  The paper is organized as follows. In section II, we briefly describe and discuss about the Yukawa
  structure of aligned 2HDM type III. Section III explains about the method of numerical analysis in this work.
  In section IV, we study the top quark FCNC processes and investigate  the bounds from the LHC experiment.
  In section V and VI, we study the constraints from the flavor physics with tree-level and loop-level processes.
  Section VII is reserved for the combined analysis and future prospect for the constraints on $\lambda_{ct}$.
  We conclude and summarize our result in section VIII.

\section{Yukawa Sector of aligned 2HDM type III}
The Yukawa interaction Lagrangian of 2HDM type III can be described as~\cite{Mahmoudi:09}
\begin{equation}
\label{eq:YukawaL}
        - \mathcal L_{\rm Y}=\bar Q_L( Y_1^d\Phi_1+Y_2^d\Phi_2) d_R+\bar Q_L(Y_1^u \tilde\Phi_1+Y_2^u \tilde\Phi_2)u_R+\bar L_L(Y_1^\ell\Phi_1+Y_2^\ell\Phi_2) e_R+h.c.,
\end{equation}
where $Q_L, L_L$ are left-handed quark and lepton doublets while
$u_R, d_R, e_R$ are right-handed singlets in interaction basis.
The two Higgs doublets $\Phi_1$ and $\Phi_2$ are introduced with
the definition $\tilde\Phi_{i}=i\sigma_2 \Phi_i^*$ where
$\sigma_2$ is Pauli matrix. $Y_{1,2}^{u,\,d,\,\ell}$ are
corresponding Yukawa matrices where the flavor indices are
implicitly considered. After the EW symmetry breaking $\Phi_1$ and
$\Phi_2$ have the vevs $\langle \Phi_i \rangle = v_i /\sqrt{2}$
which satisfies $v_1^2+v_2^2=v^2$, where $v=246\GeV$. As usual, we
define $\tan\beta = v_2/v_1$.

 Then, we diagonalize mass matrices for fermions from Eq.~(\ref{eq:YukawaL}) and for Higgses from Higgs potential
 Lagrangian which is described in many literatures
 (We refer to review paper Ref.~\cite{Branco:11}).
 We define $\alpha$ as a mixing angle of neutral CP-even Higgses.
 As we discussed in the introduction, we adopt the alignment limit that specifies
 \begin{equation}
 \sin(\beta-\alpha) = 1\,,
 \end{equation}
 to make the model comply with the Higgs precision measurement~\cite{2hdm:Higgs:fit1,2hdm:Higgs:fit2,2hdm:Higgs:fit3,2hdm:Higgs:fit4,
2hdm:Higgs:fit5,2hdm:Higgs:fit6,2hdm:Higgs:fit7,2hdm:Higgs:fit8}.
 With this alignment limit, the Yukawa Lagrangian Eq.~(\ref{eq:YukawaL}) is re-expressed in terms of mass eigenstates as follows
 \begin{eqnarray}
 \label{eq:YukawaL2}
 \mathcal L_{\rm Y} &=& \mathcal L_{\rm Y, \,SM}
 + \frac{1}{\sqrt{2}}\,\bar d \xi^d d H
 + \frac{1}{\sqrt{2}}\,\bar u \xi^u u H
  + \frac{1}{\sqrt{2}}\,\bar \ell \xi^\ell \ell H
 - \frac{i}{\sqrt{2}} \,\bar d \gamma_5 \xi^d d A
 - \frac{i}{\sqrt{2}}\, \bar u \gamma_5 \xi^u u A \nn \\
 &&  - \frac{i}{\sqrt{2}} \,\bar \ell \gamma_5 \xi^\ell \ell A
 + \Big[\bar u \Big(\xi^u V_{\rm CKM} P_L -  V_{\rm CKM} \xi^d P_R \Big)d H^+ - \bar \nu \xi^\ell P_R \ell H^+ + h.c. \Big]\,,
 \end{eqnarray}
 by ignoring Goldstone Lagrangian.
 Here, ${\cal L}_{\rm Y,\,SM}$ is equal to the SM Yukawa Lagrangian, $u,d,\ell$ are mass eigenstates of up- and down-type quarks and leptons, $H,A$ are CP-even and -odd neutral Higgses, and $H^\pm$ are charged Higgses. $V_{\rm CKM}$ is the CKM matrix, $P_L$ and $P_R$ are chiral projection operators, $P_{L,R} = \frac{1}{2}(1\mp \gamma_5)$. Note that in the alignment limit, the SM Yukawa sector is completely decoupled from the NP sector. $\xi^{u,\,d,\,\ell}$ are Yukawa matrices for the mass eigenstates which include all the FCNC couplings.

 In this work we assume that the new Yukawa matrices are CP-conserving, that is  $\xi^{u,\,d,\,\ell}$ are real and symmetric :
 \begin{equation}
 \xi^{u,\,d,\,\ell}_{ij} = \xi^{u,\,d,\,\ell\,*}_{ij}= \xi^{u,\,d,\,\ell}_{ji}\,.
 \end{equation}
 To avoid severe constraints from down-type quark FCNC, we adopt Cheng-Sher ansatz, Eq~(\ref{eq:ChengSher}). 
 Due to the tiny masses of $u,d,s$ quarks, the elements of Yukawa couplings that contain those quarks are negligibly small:
 \begin{align}\label{eq:Yuk}
  \xi^d \simeq
  \begin{pmatrix}
    0 & 0 & 0 \\
    0 & 0 & \xi_{sb} \\
    0 & \xi_{sb} & \xi_{bb}
  \end{pmatrix},
~~~
  \xi^u \simeq
  \begin{pmatrix}
    0 & 0 & 0 \\
    0 & \xi_{cc} & \xi_{ct}\\
    0 & \xi_{ct} & \xi_{tt}
  \end{pmatrix}\,.
\end{align}
Here, we include $\xi_{sb}$ since it can play some role in our study.
 In this set-up, the only relevant top-quark FCNC coupling is $\lambda_{ct}$ where $\xi_{ct} = \lambda_{ct} \sqrt{2m_c m_t}/v \,$. It should be emphasized that the top quark FCNC coupling  $\lambda_{ct}$ not only belongs to neutral Higgs Yukawa sector but also comes into play in charged Higgs Yukawa sector as can be seen in Eq.~(\ref{eq:YukawaL2}). This important feature leads us to probe $\lambda_{ct}$ with the combined analysis of phenomenologies of both collider physics via neutral Higgs exchange and flavor physics via charged Higgs exchange.

\section{Method of numerical analysis}
\begin{table}[t]
  \centering
  \begin{tabular}{l l l}
    \hline
    $|V_{us}| f_+^{K\to \pi}(0)$  & $0.21664\pm 0.00048 $ & \cite{Charles:2015gya}
    \\
    $|V_{ub}|$ (semi-leptonic) & $(3.70\pm 0.12 \pm 0.26) \times 10^{-3}$ & \cite{Charles:2015gya}
    \\
    $|V_{cb}|$ (semi-leptonic) & $(41.0 \pm 0.33 \pm 0.74) \times 10^{-3}$ & \cite{Charles:2015gya}
    \\
    $\gamma [^\circ]$ & $73.2_{-7.0}^{+6.3}$ & \cite{Charles:2015gya}
    \\\hline
    $\overline{m}_c (\overline{m}_c)$ & $(1.286 \pm 0.013 \pm 0.040) \, {\rm GeV}$ & \cite{Charles:2015gya}
    \\
    $\overline{m}_b (\overline{m}_b)$ & $(4.18 \pm 0.03 ) \, {\rm GeV}$ & \cite{PDG}
    \\
    $\overline{m}_t (\overline{m}_t)$ & $(165.95 \pm 0.35 \pm 0.64) \, {\rm GeV}$ & \cite{Charles:2015gya}
    \\\hline
    $f_+^{K\to \pi}(0)$ & $0.9641 \pm 0.0015 \pm 0.0045$& \cite{Charles:2015gya}
    \\
    $f_{B_s}$               & $(225.6 \pm 1.1 \pm 5.4)\,{\rm MeV}$ & \cite{Charles:2015gya}
    \\
    $f_{B_s}/f_{B_d}$     & $1.205 \pm 0.004 \pm 0.007$   &\cite{Charles:2015gya}
    \\
    $\hat{B}_{B_s}$       & $1.320\pm 0.017 \pm 0.030$ & \cite{Charles:2015gya}
    \\
    $\hat{B}_{B_s}/\hat{B}_{B_d}$ & $1.023 \pm 0.013 \pm 0.014$ & \cite{Charles:2015gya}
    \\\hline
  \end{tabular}
  \caption{\baselineskip 3.0ex
  The theoretical input parameters used in the numerical analysis.}
  \label{tab:input}
\end{table}

 Before we discuss the phenomenology of top quark FCNC, we first summarize theoretical input parameters as well as experimental values that are used in this work and discuss about the details of numerical analysis.
Table~\ref{tab:input} shows input parameters for the processes in flavor physics. The values are taken from the latest result of CKMfitter collaboration~\cite{CKMfitter}.
To obtain the uncertainties of theory prediction, we vary each parameter value within $1\sigma$ range and add each individual uncertainty in quadrature.

In Table~\ref{tab:exp} we summarize experimental data and their  SM predictions by using the input values in Table~\ref{tab:input}.  We note that all the SM predictions are in good agreement with the current experimental data, except the ratio $R(D^*)$ which will be discussed in later section. For each observable, the relevant parameters for the theory prediction in 2HDM type III are enumerated. Apparently, those parameters will be constrained by corresponding
experimental data. The detailed discussions are presented in the following sections.

As discussed in the previous sections, the relevant model
parameters we are interested in aligned 2HDM type  III include three mass parameters
$M_{H^\pm}$, $M_H$, $M_A$, and four Yukawa couplings
$\lambda_{\tau\tau}$, $\lambda_{bb}$, $\lambda_{tt}$, and
$\lambda_{ct}$. Here, we  choose the light neutral Higgs boson $h$
as the observed Higgs boson at the LHC and adopt the alignment
limit~\cite{Craig:13,Celis:13,Cheng:14,Song:2014lua}. For other choice
that the heavy neutral Higgs $H$ is observed one, we refer to
Ref.~\cite{Bernon:2014nxa,Chang:2015goa}. Direct searches for
charged Higgs bosons have been performed at LEP~\cite{ALEPH:02},
Tevatron~\cite{CDF:05,D0:09} and LHC~\cite{CMS:12,ATLAS:13b}. The
LEP Collaboration put the lower bound $M_{H^\pm} \geq 79.3 {\,\rm
GeV}$ by assuming $\mathcal B(H^+ \to \tau^+ \nu)+ \mathcal B( H^+
\to  c \overline s)=1$ within 2HDM~\cite{ALEPH:02}. The neutral Higgs
search at the LEP experiment also put lower bound on the neutral
Higgs masses such as $M_H>92.8\GeV$ and $M_A>93.4\GeV$ within
CP-conserving MSSM scenario~\cite{Schael:2006cr}. We adopt those
lower limits for heavy Higgs masses as reference values even
though above results may depend on Yukawa structure and $m_{\rm
SUSY}$ scale. Indeed, the lower limits of Higgs masses are
irrelevant to our main result.
With all these considerations, we restrict the
parameters of 2HDM type III in the following ranges:
\begin{align}
\label{eq:paramscan}
M_{H^\pm} &\in [\,80,1000]\GeV\,, \nn\\
M_H\,(M_A) &\in [\,125\,(93), 1000]\GeV\,.
\end{align}
These choices of parameter regions are shown to be reasonable in later section.

\begin{table}[t]
  \centering
  \begin{tabular}{l l l l l}
    \hline
    observable & SM & EXP & Ref & 2HDM parameters\\
    \hline
    $\mathcal B(B \to \tau \nu)\cdot 10^{4}$ & $0.85\pm 0.14$& $1.14 \pm 0.22$ & \cite{HFAG} & 
    $\lambda_{bb}$, $\lambda_{bs}$, $\lambda_{bd}$, $\lambda_{ut}$,
    $\lambda_{\tau\tau}$, $M_{H^\pm}$
    \\
    $R(D)$ &$0.297\pm 0.017$& $0.391 \pm 0.041 \pm 0.028$ & \cite{HFAG:EPS15} & \tg{$\lambda_{bb}$},
    $\lambda_{\tau\tau}$, $\lambda_{ct}$,  $M_{H^\pm}$
    \\
    $R(D^*)$ & $0.252 \pm 0.003$& $0.322\pm 0.018\pm0.012$ & \cite{HFAG:EPS15} & \tg{$\lambda_{bb}$}, $\lambda_{\tau\tau}$, $\lambda_{ct}$, $M_{H^\pm}$
    \\\hline
    $\Delta m_d [\,{\rm ps}^{-1}]$ & $0.51 \pm 0.06 $& $0.510 \pm 0.003 $ & \cite{HFAG} & \tg{$\lambda_{bb}$},  $\lambda_{tt}$, $\lambda_{ct}$,  $M_{H^\pm}$
    \\
    $\Delta m_s [\,{\rm ps}^{-1}]$ &  $16.93 \pm 1.16$ & $17.757 \pm 0.021 $ & \cite{HFAG} & \tg{$\lambda_{bb}$},  $\lambda_{bs}$,
     $\lambda_{tt}$, $\lambda_{ct}$, $M_{H^\pm}$
    \\
    $\mathcal B(B\to X_s \gamma) \cdot 10^{4}$ & $3.36 \pm 0.23$& $3.43\pm 0.22$&\cite{HFAG} & $\lambda_{bb}$, $\lambda_{tt}$, $\lambda_{ct}$,  $M_{H^\pm}$
    \\\hline
    $\mathcal B(t \to c g)$& $<10^{-10}$ & $<1.6 \times 10^{-4}$ (95\% CL)& \cite{ATLAS:13} & \tg{$\lambda_{bb}$}, $\lambda_{tt}$, $\lambda_{ct}$, \tg{$M_{H^\pm}$}, $M_{H}$, $M_A$
    \\
    $\sigma(pp \to tt)$ &-& $<62 {\,\rm fb}$  (95\% CL)&\cite{Aad:2015gdg}& $\lambda_{ct}$, $M_H$, $M_A$
    \\
    $R_b$ & $0.21576\pm 0.00003$ & $0.21629\pm 0.00066$& \cite{EWPT:09}& \tg{$\lambda_{bb}$}, $\lambda_{tt}$, $\lambda_{ct}$, $M_{H^\pm}$
    \\
    $\rho_0$ &$1$& $1.00040\pm0.00024$& \cite{PDG}& $M_{H^\pm}$, $M_H$, $M_A$
    \\\hline
  \end{tabular}
  \caption{\baselineskip 3.0ex
  SM predictions and experimental measurements for the observables used in the numerical analysis. The last column denotes their dependence on the 2HDM parameters. The parameters in the parenthesis imply that they can be safely neglected.
  }
  \label{tab:exp}
\end{table}

In order to derive an allowed parameter space, we impose the experimental constraints in the same way as in Refs.~\cite{Cheng:14,Jung:12}: for each point in the theoretical parameter space we span the range of the theory prediction for an observable by performing the $2\sigma$ variations of input parameters.
If the difference between the central values of theory prediction and experimental value is less then the sum of two errors in quadrature, then
 this point is regarded as allowed. Since the main theoretical uncertainties are due to the hadronic input parameters, common to both the SM and the 2HDM, the relative theoretical uncertainty is assumed to be constant at each point in the parameter space.

\section{top quark FCNC processes at colliders}
The LHC is often called top-factory since the top pair is copiously produced through QCD interaction. The LHC Run I data already collected millions of top pair events, and even much more top pair events are expected to be collected in the LHC Run II. Undoubtedly, the LHC provides us unique chance to explore the top quark FCNC processes which are extremely small in the SM.

The experimental search for top quark FCNC can be performed either by anomalous decays or production of top quarks at hadron colliders with top quark FCNC couplings~\cite{Heng:2009wr,Lu:2010zzb,Khatibi:2014via,Greljo:2014dka,Wu:2014dba,Hesari:2015oya}.
We note that the searches for $t\to ch$~\cite{CMS:2014topfcnc1,Aad:2014dya} do not provide any constraints on 2HDM type III in alignment limit since the top quark FCNC couplings with the SM Higgs vanish. The anomalous top decays via $t \to c/u \,V$ where $V=\gamma,Z$ are explored at the Tevatron~
\cite{Abe:1997fz,Aaltonen:2008ac,Abazov:2011qf} and at the LHC~\cite{Aad:2012ij,Chatrchyan:2012hqa,Chatrchyan:2013nwa,CMS:2014topfcnc2}, without finding any significant excess of signal events.
However, these searches do not provide any meaningful constraints on 2HDM type III since the prediction is much suppressed by loop correction and EW couplings.
 Contrary to top decays, the anomalous single top production has much chance to probe top quark FCNC coupling due to the large gluon luminosity in the parton-distribution-function (PDF) and the relatively large QCD coupling. The experimental searches for single top events put upper bound on $\Br(t\to cg)$ and $\Br(t\to ug)$~\cite{Aaltonen:2008qr,Abazov:2010qk,ATLAS:2013topfcnc1,CMS:2014topfcnc3,CMS:2014topfcnc4}.
 We  focus on $\Br(t\to cg)$ by ignoring $u$-quark involved FCNC process since it is extremely suppressed in Cheng-Sher ansatz even though $u$ quark PDF is bigger than $c$ quark PDF.

The same sign top pair production is a tree-level process and
therefore promising to test NP scenarios which contain top quark
FCNC couplings. Notable example is that the NP scenario with
$Z^\prime$ mediated top quark FCNC coupling~\cite{Jung:2009jz,Cao:2011ew}
that explains the anomalous top forward-backward asymmetry
observed at the
Tevatron~\cite{Abazov:2007ab,Aaltonen:2008hc,Aaltonen:2011kc} is
disfavored by non-observation of the same sign top pair production
at the LHC~\cite{Chatrchyan:2011dk,Aad:2012bb}. The recent
experimental search at ATLAS with integrated luminosity of
$20.3\fb^{-1}$ at $8\TeV$ puts the most stringent upper limits on
$\sigma(pp\to tt)$. We interpret the result as an upper limit on
$cc\to tt$ process to constrain $\lambda_{ct}$.

In what follows, we study the phenomenology of $t\to cg$ and $cc\to tt$ processes within the 2HDM type III to investigate the top quark FCNC coupling.

\subsection{$t\to cg$}

In the SM, $t\to cg$ decay is extremely suppressed due to GIM mechanism. However, this rare top decay can be enhanced in some NP scenarios~\cite{delAguila:98,Yang:97}. In general, the form factor for the effective $tcg$ vertex is defined by~\cite{Atwood:96}
\footnote{\baselineskip 3.0ex
In Ref.~\cite{Atwood:96}, the last two terms of Eq.~(\ref{eq:formfactor}) are omitted. Although they do not contribute to the width $\Gamma(t\to cg)$, they are necessary  to satisfy Ward identity.}
\begin{align}\label{eq:formfactor}
  \mathcal L^{\rm ctg} =\frac{1}{16\pi^2} \bar c \left( {\cal A} \gamma^\mu + {\cal B} \gamma^\mu \gamma_5 +i {\cal C} \sigma^{\mu\nu}\frac{q_\nu}{m_t} +i {\cal D} \sigma^{\mu\nu}\frac{q_\nu}{m_t}\gamma_5 - {\cal A} \frac{m_t}{q^2}q^\mu +{\cal B} \frac{m_t}{q^2} \gamma_5 q^\mu \right)t G_\mu^a T^a,
\end{align}
where $T^a$ ($a=1,\dotsc,8$) denote $SU(3)$ generators. The form factors ${\cal A}$, ${\cal B}$, ${\cal C}$ and ${\cal D}$ have been calculated in various types of 2HDM~\cite{Eilam:1990zc,Grzadkowski:90,Abbas:15}. In the 2HDM type III, these form factors are generated by the penguin diagrams mediated by the neutral Higgses $h$, $H$ and  $A$ and  charged Higgs $H^\pm$. Their explicit expressions are given in Appendix~\ref{sec:FF}.
\begin{figure}[t]
  \centering
\includegraphics[width=0.6\textwidth]{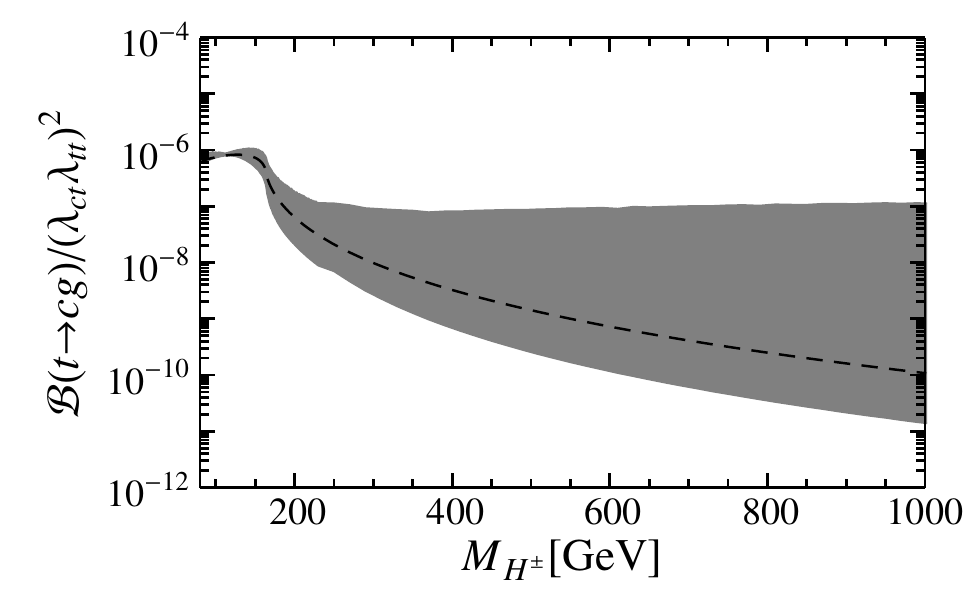}
  \caption{\baselineskip 3.0ex
  Branching ratio of $t \to cg$ as a function of the charged Higgs mass. Dashed line: a common scalar mass $M_{H^\pm}=M_H=M_A$ is taken. Shaded region: neutral Higges' masses $M_H$ and $M_A$ vary but constrained by the oblique parameter $\Delta\rho$.}
  \label{fig:BR_t_c_g}
\end{figure}
With the convention Eq.~(\ref{eq:formfactor}), the decay width for $t \to c g$ is given by~\cite{Atwood:96}
\begin{align}
  \Gamma (t \to cg)=\frac{1}{(16\pi^2)^2}\frac{1}{8\pi}m_t C_F (|{\cal C}|^2+|{\cal D}|^2),
\end{align}
with $C_F=(N_c^2-1)/2N_c$. We note that  $\Br(t \to cg)$ is proportional to $(\lamct\lamtt)^2$ as can be seen from Eq.~(\ref{eq:xidef}).

The LHC search for anomalous single top production is performed by ATLAS Collaboration with $14.2\invfb$ at $8\TeV$~\cite{ATLAS:13}. Non-observation of signal put an upper limit on $\Br(t \to cg)$ as
\begin{equation}
\Br(t \to cg) < 1.6\times 10^{-4}\,.
\end{equation}
In Fig.~\ref{fig:BR_t_c_g} we show the plot of 2HDM type III prediction for $\Br (t\to cg)$ as a function of the charged Higgs mass by setting $\lamct\lamtt=1$.\footnote{\baselineskip 3.0ex
Our numerical result is consistent with Fig. 3 of Ref.~\cite{Atwood:96} by setting  $\xi_{ct}=\xi_{tt}=1$.
} The shaded region is spanned by changing neutral Higgses masses under the constraints from $\Delta \rho$. We refer to Ref.~\cite{Song:2014lua} for detailed analysis of $\Delta \rho$. Even though there can be up to factor ${\cal O}\,(10^3)$ enhancement comparing to the SM expectation for the small $M_{H^\pm}$, the current experimental bound is far above the theory prediction. Therefore, it would be hard to constrain the top quark FCNC parameter space with anomalous single top production measurement at the LHC.

\subsection{$cc \to tt$}
The same sign top pair production at hadron collider requires FCNC
coupling with $t-$ or $u-$channel exchange of neutral particle
with spin $0$ or $1$ since the electric charges of final states
are same. Another possibility is $s$-channel process mediated by
a charge 4/3 new particle. Various NP scenarios that
contribute to the same sign top pair production are well
summarized in Ref.~\cite{AguilarSaavedra:2011zy} with effective
operator formalism. The production rate of the same sign top pair
at hadron colliders via the contact interactions with different
chiral configuration is modeled in
Ref.~\cite{AguilarSaavedra:2010zi}. Meanwhile, in this work we
perform the full theory analysis with spin $0$ Higgs boson as a
mediator since the effective operator formalism may not reproduce
well the full theory result if the mediator mass is quite less
than $1\TeV$. We refer to Ref.~\cite{Goldouzian:2014nha} for the analysis with another mediators.

In the 2HDM type III with alignment limit, the same sign top pair production arises at tree level via $t$- or $u$-channel diagrams with exchange of heavy neutral Higgs bosons, $H$ or $A$. The partonic scattering cross section for $qq\to tt$ process is described as
\begin{equation}
{\hat \sigma}(\hat s) = \int d{\hat t} \frac{1}{64 \pi {\hat s}^2 N_c}\Big( {\hat g}_H({\hat s},{\hat t})+{\hat g}_A({\hat s},{\hat t})+{\hat g}_{\rm intf}({\hat s},{\hat t}) \Big)\,,
\end{equation}
where the amplitude square functions ${\hat g}_i$ are defined as
\begin{eqnarray}
{\hat g}_\phi ({\hat s},{\hat t}) &=& N_c^2 \xi_{ct}^4 \bigg[ \bigg(\frac{t-m_t^2}{t-M_\phi^2}\bigg)^2+ \bigg(\frac{u-m_t^2}{u-M_\phi^2}\bigg)^2+\frac{tu-m_t^2s-m_t^4}{N_c(t-M_\phi^2)(u-M_\phi^2)}\bigg]
\,, \nn \\
{\hat g}_{\rm intf}  ({\hat s},{\hat t}) &=&
2N_c \xi_{ct}^4 \frac{\big(tu+m_t^2s-m_t^4\big)\big(tu+(M_H^2+M_A^2)(s/2-m_t^2)+M_H^2 M_A^2\big)}
{(t-M_H^2)(t-M_A^2)(u-M_H^2)(u-M_A^2)}\,,
\end{eqnarray}
where $\phi =H,A$. Then the total cross section for $cc\to tt$ is convoluted with parton luminosity function $f_{cc}(x,\mu_F)$ of sea quark pair $cc$ as follows
\begin{equation}
\sigma(cc\to tt) = \int_\tau^1 dx {\hat \sigma}(xs) f_{cc}(x,\mu_F)\,,
\end{equation}
where $\tau = 4m_t^2/s$ and $f_{cc}(x,\mu_F)$ is defined by
\begin{equation}
f_{cc}(x,\mu_F) = \int_x^1 \frac{dy}{y} f_{c/p}(y,\mu_F) f_{c/p}(x/y,\mu_F)\,.
\end{equation}
Here, $f_{c/p}(y,\mu_F)$ is $c$-quark PDF and the factorization scale $\mu_F$ is set to be $\mu_F = m_t$. We use MSTW2008LO PDF set~\cite{Martin:2009iq} for the numerical analysis. The gluon and charm quark initial state process with extra jet radiation is not considered by assuming that the contribution is subleading.

\begin{figure}[t]
  \centering
    \subfigure{\label{fig:cc2tt1}\includegraphics[width=0.465\textwidth]{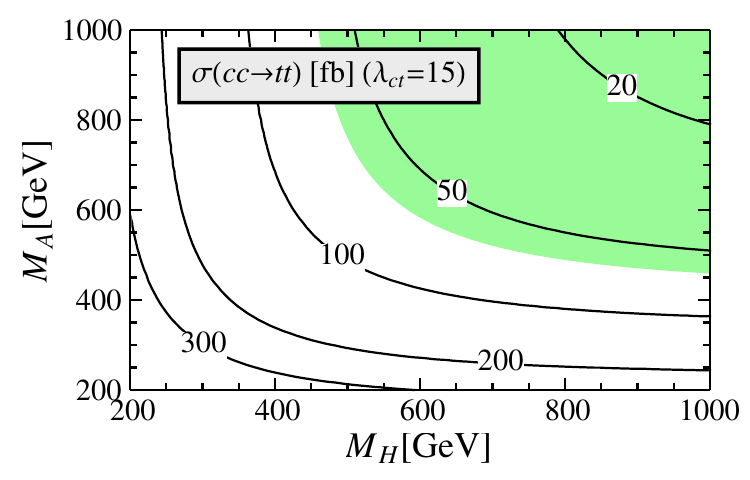}}
  \qquad
  \subfigure{\label{fig:cc2tt2}\includegraphics[width=0.465\textwidth]{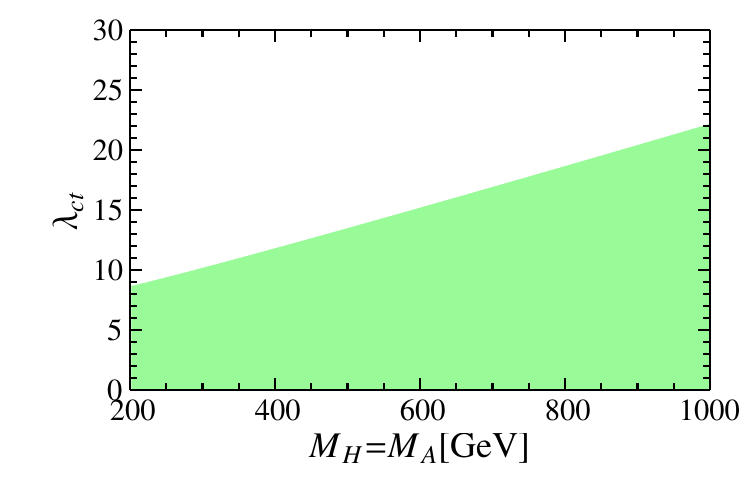}}
  \caption{\baselineskip 3.0ex
  (a) Total cross section for $cc\to tt$ at the LHC 8TeV run in $(M_H,M_A)$ plane. We set $\lambda_{ct}=15$. The shaded region (green) is allowed parameter space at 95\%\,CL. (b) The allowed parameter space in $(M_H(=M_A),\lambda_{ct})$ plane in the case where $H$ and $A$ are degenerated in mass.
  }
  \label{fig:cc2tt}
\end{figure}

The experimental searches for the same-sign dileptons and $b$-jets
at CMS with $19.5\invfb$~\cite{Chatrchyan:2013fea} and at ATLAS
with $20.3\invfb$~\cite{Aad:2015gdg} at $8\TeV$ can be applied for
constraining the same-sign top pair production rate. The
non-observation of any significant excess of signal events sets
the upper bound of the production cross section. The strongest
bound comes from ATLAS result. ATLAS provides different upper
bounds depending on the helicity configuration of effective
operators within contact interaction model. We conservatively
adopt the largest upper bound among the three as follows:
\begin{equation}
\sigma(pp\to tt) < 62\fb~~(\textrm{ATLAS 95\%\,CL~\cite{Aad:2015gdg}}) \,.
\end{equation}

We re-interpret this result to constrain the cross section  $\sigma(cc\to tt)$ using the formula described above.
The constraint is usually strong for small Higgs masses. Since the signal rate is proportional to $\lambda_{ct}^4$, the large values of $\lambda_{ct}$ are severely constrained and conversely the small value of $\lambda_{ct}$ is hardly excluded. Fig.~\ref{fig:cc2tt1} shows the prediction of scattering cross section by setting $\lambda_{ct} = 15$ in $(M_H,M_A)$ plane and the allowed region with shaded green color. As shown, the interference effect is constructive. For the given $\lambda_{ct}$ value the region $M_H, M_A \lesssim 400\GeV$ is excluded. Fig.~\ref{fig:cc2tt2} shows the allowed parameter space in $(M_H,\lambda_{ct})$ plane for the case where $H$ and $A$ are degenerated in mass. Experimental bound provides quite stringent upper limit on $\lambda_{ct}$ as $10\sim20$, depending on the heavy Higgs mass.

\section{Flavor physics - Tree-level processes}
Since the top-quark FCNC couplings take part in charged Higgs Yukawa sector, they can  contribute to the semi-leptonic decay and leptonic decay of $B$ mesons
which are tree-level processes. In this section we study the two $\tau$-involved tree-level processes, $B\to D^{(*)}\tau \nu $ and $B\to \tau \nu$ to constrain top quark FCNC couplings. The former (latter) is involved with $b\to c(u)$ charged current. Therefore, any NP model which contains such charged current with a new charged particle can contribute to these processes~\cite{Ko:2012hd,Ko:2012sv,Freytsis:2015qca,Cvetic:2015roa}.

For those processes with $b\to c(u)$ charged current, the effective Hamiltonian is described by~\cite{Crivellin:12}
\begin{align}\label{eq:Hamiltonian:BDtaunu}
  \mathcal H_{\rm eff}=C^{\,q}_{\rm VLL}{\mathcal O}^{\,q}_{\rm VLL}+C^{\,q}_{\rm SRL}{\mathcal O}^{\,q}_{\rm SRL}+C^{\,q}_{\rm SLL} {\mathcal O}^{\,q}_{\rm SLL},~~~(q=u,c)
\end{align}
with the effective four-fermion operators
\begin{align}
  {\mathcal O}^{\,q}_{\rm VLL}&=(\bar q \gamma_\mu P_L b)(\bar \tau \gamma^\mu P_L \nu_\tau),\nonumber\\
  {\mathcal O}^{\,q}_{\rm SRL}&=(\bar q P_R b)( \bar\tau P_L \nu_\tau),\nonumber\\
  {\mathcal O}^{\,q}_{\rm SLL}&=(\bar q P_L b)( \bar\tau P_L \nu_\tau)\,.
\end{align}
Within the SM, the vector boson $W^-$ is exchanged, therefore only ${\mathcal O}^{\,q}_{\rm VLL}$ are generated with tree-level Wilson coefficients
\begin{equation}
\label{eq:WC1}
 C_{\rm VLL}^{\,q,\,\rm SM}=\frac{4G_F V_{qb}}{\sqrt 2}\,,
\end{equation}
where $G_F$ denotes the Fermi coupling constant and $V_{qb}$ are the CKM matrix elements. On the other hand, within the 2HDM type III the scalar charged Higgs boson is exchanged, and therefore ${\mathcal O}^{\,q}_{\rm SLL}$ and ${\mathcal O}^{\,q}_{\rm SRL}$ are generated. The corresponding tree-level Wilson coefficients are
\begin{eqnarray}
\label{eq:WC2}
  C_{\rm SLL}^{\,c,\,\rm 2HDM}=\frac{V_{tb} \xi_{ct} \xi_{\tau\tau}}{M_{H^{\pm}}^2}\,,~~~
    C_{\rm SRL}^{\,q,\,\rm 2HDM}=-\frac{V_{qb} \xi_{bb} \xi_{\tau\tau}}{M_{H^{\pm}}^2}\,.
\end{eqnarray}
We neglect $C_{\rm SLL}^{\,u,\,\rm 2HDM}$ which is proportional to $\lambda_{ut}$ and extremely suppressed by $u$-quark mass.

For $B \to D^{(*)}\tau\nu$ decay, we can define a theoretically clean observable by taking the ratio with relatively clean measurement $B \to D^{(*)}\ell\nu\,(\ell=e,\mu,\tau)$ to cancel the hadronic uncertainties:
\begin{align}
  R(D^{(*)}) \equiv\frac{\mathcal B (B \to D^{(*)}  \tau\nu)}{\mathcal B(B\to D^{(*)}\ell\nu)}\,.
\end{align}
Note that the CKM matrix element $V_{cb}$ is also canceled out. Then, the theory uncertainty of $ R(D^{(*)})$ are very small, $6(1)\%$, while the experimental error is quite large, $12(7)\%$ because of missing neutrino in $\tau$ reconstruction.

 With the effective Hamiltonian in Eq.~(\ref{eq:Hamiltonian:BDtaunu}), the theoretical prediction of $ R(D^{(*)})$ relative to the SM value is described as~\cite{Akeroyd:03,Fajfer:12,Sakaki:13,Crivellin:12},
\begin{align}
\label{eq:RDexpr}
  R(D)  &=R_{\rm SM}(D)\left (1+1.5{\rm Re}\left [\frac{C^{\,c}_{\rm SRL}+C^{\,c}_{\rm SLL}}{C_{\rm VLL}^{\,c,\,\rm SM}} \right] + 1.0 \left\lvert \frac{C^{\,c}_{\rm SRL}+C^{\,c}_{\rm SLL}}{C_{\rm VLL}^{\,c,\,\rm SM}} \right\rvert^2 \right),\nonumber\\
  R(D^*)  &=R_{\rm SM}(D^*)\left (1+0.12{\rm Re}\left [\frac{C^{\,c}_{\rm SRL}-C^{\,c}_{\rm SLL}}{C_{\rm VLL}^{\,c,\,\rm SM}} \right] + 0.05 \left\lvert \frac{C^{\,c}_{\rm SRL}-C^{\,c}_{\rm SLL}}{C_{\rm VLL}^{\,c,\,\rm SM}} \right\rvert^2 \right).
\end{align}
Due to the spin of $D^*$ meson, the NP effects on $R(D^*)$ are much smaller than the ones on $R(D)$~\cite{Fajfer:12,Tanaka:94,Itoh:04,Nierste:08}. The relevant Wilson coefficients are given in Eqs.~(\ref{eq:WC1}) and (\ref{eq:WC2}).
Since
$C_{\rm SRL}^{\,c}$ is suppressed by $m_b/v$ in Cheng-Sher ansatz and also by CKM matrix element, its contribution is negligibly small.

The BaBar experimental data for $B \to D^{(*)}\tau\nu$ have shown somewhat large values comparing with the SM expectations for both $R(D)$ and $R(D^*)$ where the combined discrepancy was $3.4\sigma$ level~\cite{Lees:2012xj,Lees:2013uzd}.  It was also discussed that these can not be simultaneously accommodated  by 2HDM Type II.
To explain both discrepancies it was shown that the large top quark FCNC coupling $\lambda_{ct}$ which contributes to $C^{\,c}_{\rm SLL}$ in Eq.~(\ref{eq:RDexpr}) is needed~\cite{Chen:13,Crivellin:12}. Very recently, the Belle collaboration reported the measurements of both $R(D)$ and $R(D^*)$~\cite{Huschle:2015rga}, and the LHCb collaboration did for $R(D^*)$~\cite{Aaij:2015yra}. Even though the Belle result is in the middle of the SM expectation and the BaBar result, due to the reduced errors, the average values are still in $3.9\sigma$
discrepancy~\cite{HFAG:EPS15} (See Table 2 for comparison).

\begin{figure}[t]
  \centering
\includegraphics[width=0.6\textwidth]{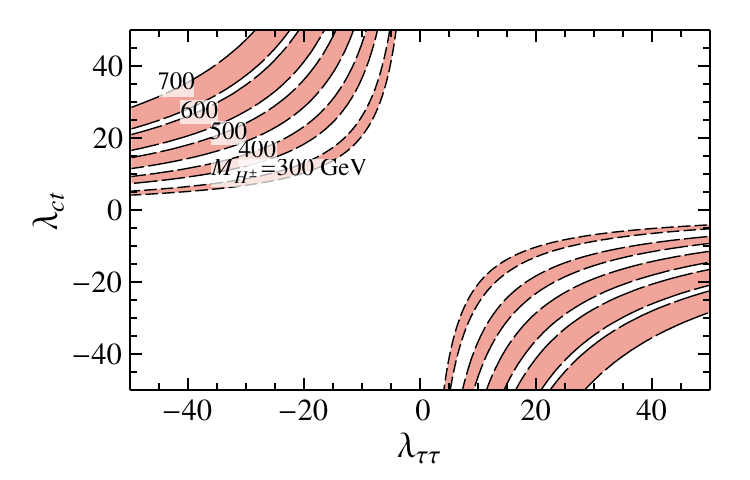}
 \caption{\baselineskip 3.0ex
  Constraints on $\lambda_{ct}$ and $\lambda_{\tau\tau}$ from $R(D)$ and $R(D^*)$. Pink-colored region is allowed at 95\% CL in ($\lambda_{\tau\tau}$, $\lambda_{ct}$) plane with different charged Higgs masses.}
  \label{fig:treelevel}
\end{figure}

The allowed parameter space in $(\lambda_{\tau\tau},\lambda_{ct})$ with different charged Higgs masses constrained by $R(D^{(*)})$ is shown in Fig.~\ref{fig:treelevel}. For any given charged Higgs mass both $\lambda_{ct}$ and $\lambda_{\tau\tau}$ do not simultaneously become zero. For small $\lambda_{\tau\tau}$ value,   $\lambda_{ct}$ must be very large.
Interestingly, larger charged Higgs mass requires larger $\lambda_{ct}$.
These feature can be understood as a whole since only the product $\lambda_{ct}\lambda_{\tau\tau}/M_{H^\pm}^2$ enters the contributions from 2HDM, as show in Eq.~(\ref{eq:WC2}). Explicitly, the current $B \to D^{(*)}\tau\nu$ data put the bound
\begin{align}
\label{eq:treebound}
  -0.0030<\lambda_{ct}\lambda_{\tau\tau}/M_{H^\pm}^2<-0.0023,
\end{align}
which can be seen in Fig.~\ref{fig:treelevel}. It is noted
that $\lambda_{\tau\tau}$ is associated with the neutral Higgs
decay $H/A \to \tau\tau$. If $\lambda_{\tau\tau}$ is large, the
LHC has a good opportunity to detect neutral Higgs bosons in their
tauonic decay channels. In the case of small $\lambda_{\tau\tau}$,
the coupling $\lambda_{ct}$ should be large, which may be severely
constrained by the same sign top pair production as shown in
previous section.

Contrary to $B \to D^{(*)}\tau\nu$ decay, $B \to \tau\nu$ decay is a helicity suppressed process and more strongly suppressed by CKM factor. Therefore, $B \to (\mu/e)\nu$ decays are extremely rare, ${\cal O}(10^{-7})$ and ${\cal O}(10^{-11})$ respectively, and not yet measured although $B\to \mu \nu$ will be measured soon at Belle\;II. Thus, we have no way to cancel the large theory uncertainty of hadronic current of $B \to \tau\nu$.
The uncertainties from the SM prediction and experiment for $\Br(B \to \tau\nu)$ are very large, 24\% and 19\% respectively.
Due to these large errors, the constraint from  $\Br(B \to \tau\nu)$ is not much significant.

With the effective hamiltonian in Eq.~(\ref{eq:Hamiltonian:BDtaunu}), the branching ratio of $B \to \tau\nu$ reads~\cite{Crivellin:12}
\begin{equation}
  \mathcal B (B \to \tau\nu) =\frac{G_F^2 |V_{ub}|^2}{8\pi}m_\tau^2 m_B \tau_B  f_B^2\left( 1-\frac{m_\tau^2}{m_B^2} \right)^2  \left\lvert 1+\frac{m_B^2}{m_b m_\tau} \frac{C^{\,u}_{\rm SRL}-C^{\,u}_{\rm SLL}}{C_{\rm VLL}^{\,u,\,\rm SM}} \right\rvert^2,
\end{equation}
where $f_B$ denotes the $B$-meson decay constant. The relevant Wilson coefficients for 2HDM type III are shown in Eqs.~(\ref{eq:WC1}) and (\ref{eq:WC2}).
We note that not only $\xi_{bb}$ but also $\xi_{bs}, \xi_{bd}$  can contribute to $C_{\rm SRL}^u$ within Cheng-Sher ansatz due to the relatively large CKM factors. Even $\xi_{ut}$ can significantly contribute to $C_{\rm SLL}^u$. Due to the combination of these contributions to a single observable  $\Br(B\to \tau \nu)$, none of these Yukawa couplings get any meaningful constraints.


\section{Flavor physics - Loop-level processes}
\subsection{$B_d\to X_s \gamma$}
\label{sec:B_Xs_gm}

As for the loop-induced process we first consider $B_d\to X_s \gamma$ decay. Taking the normalization with $\Br(B_d \to X_c e {\overline \nu}_e)$, the dominant theoretical uncertainties from $m_b^5$ and CKM factor are canceled out. The effective Hamiltonian for the $B_d \to X_s \gamma$ decay read~\cite{Buchalla:95,Gambino:01}
\begin{align}
\mathcal H_{\rm eff}=-\frac{4G_F}{\sqrt 2}V_{ts}^* V_{tb} \sum_{i=1}^8 C_i \mathcal O_i,
\end{align}
where the explicit expressions of the tree or penguin operators $\mathcal O_{1-6}$ can be found in Ref.~\cite{Buras:11}. The magnetic penguin operators, $\mathcal O_7$ and $\mathcal O_8$,   which are characteristic for this decay, are defined as
\begin{align}
  \mathcal O_7=\frac{e}{8\pi^2}m_b \bar s_\alpha \sigma^{\mu\nu} (1+\gamma_5) b_\alpha F_{\mu\nu},
\qquad
   \mathcal O_8=\frac{g_s}{8\pi^2}m_b \bar s_\alpha \sigma^{\mu\nu} (1+\gamma_5) T_{\alpha\beta}^a b_\beta G_{\mu\nu}^a,
\end{align}
where $m_b$ denotes the $b$-quark mass in the $\overline{\rm MS}$ scheme, and $e$ ($g_s$) is the electromagnetic (strong) coupling constant. The heavy degrees of freedom from the $W^-$ boson contribution~\cite{bsr:SM,ali93,misiak93,buras94,cella94,ciuchini94,Adel94,Greub97,Bobeth00} and charged Higgs contribution~\cite{Ciuchini:1997xe,Borzumati:1998tg,Hermann:2012fc} are integrated out at $m_W$ scale, and we obtain the Wilson coefficients $C_{7,8}(\mu=m_W)$. They evolve into $\mu=m_b$ scale by renormalization group equation and consequently resum the large logarithms in perturbative QCD to all order~\cite{bsr:3loop:RG:Wilson,Chetyrkin98,Gambino03}. The higher order correction at $\mu=m_b$ scale should be necessarily done~
\cite{bsr:2loop:matrix:mb,Greub96,Buras01b,Buras02b}.

The compilation of all those calculation for $\Br(B_d\to X_s \gamma)$ reached at next-to-next-to-leading-order (NNLO) in perturbative QCD~\cite{Misiak:06a,Misiak:06b,Benzke:10}. (For a recent review, we refer to Ref.~\cite{Misiak:15}.) For given NP contributions to $C_{7,8}^{\rm NP}$, the theory prediction for $\Br(B_d\to X_s \gamma)$ at NNLO is given by ~\cite{Misiak:15}
\begin{align}
\label{eq:B2XgammaNNLO}
\mathcal B (B_d\to X_s\gamma)\times 10^{4}=(3.36\pm 0.23)-8.22 {\rm Re} C_7^{\rm NP}-1.99 {\rm Re} C_8^{\rm NP}\,,
\end{align}
where the first number represents the most up-to-date SM prediction.
By using current experimental data, we obtain
 \begin{equation}
 \label{eq:WCconstraints}
8.22 {\rm Re} C_7^{\rm NP}+1.99 {\rm Re} C_8^{\rm NP} = -0.07 \pm 0.32\,.
 \end{equation}
 Therefore, it is natural for $C_{7,8}^{\rm 2HDM}$ to become ${\cal O}(0.1)$.

In the 2HDM type III, the one-loop contribution to $C_{7,8}$ via charged Higgs exchange is described by ~\cite{Ciuchini:1997xe}
  \begin{align}\label{eq:BXsgamma:WC2HDM}
    C_{7,8}^{\rm 2HDM}=\frac{1}{3} A_u^* F_{7,8}^{(1)}(x_W)- A_d^*F_{7,8}^{(2)}(x_W)\,,
  \end{align}
where the loop functions $F_{7,8}^{(1,2)}$ are given in Ref.~\cite{Ciuchini:1997xe} and $x_W=m_t^2/m_W^2$.
The Yukawa components $A_u$ and $A_d$ normalized by SM ones are defined as
  \begin{align}\label{eq:lambdas}
    A_u&=\left(\lambda_{tt}+\frac{V_{cs}}{V_{ts}} \sqrt{\frac{m_c}{m_t}} \lambda_{ct}\right) \left(\lambda_{tt}+ \frac{V_{cb}^*}{V_{tb}^*} \sqrt{\frac{m_c}{m_t}}\lambda_{ct} \right),
    \,\\ \nn
    A_d&=\left( \lambda_{tt}+\frac{V_{cs}}{V_{ts}} \sqrt{\frac{m_c}{m_t}}\lambda_{ct} \right) \lambda_{bb}\,.
  \end{align}
  It should be emphasized that the $A_d$ term is enhanced by the spin-flip factor $m_t/m_b$ and becomes comparable to $A_u$. Therefore, it is unique for  $B_d \to X_s \gamma$ that  the coupling $\lambda_{bb}$ can be significantly constrained.
  Another interesting feature is that the coefficient $\lambda_{ct}$ of second factor in $A_u$ is highly suppressed while the one in first term contains CKM-enhanced factor.
  The $\lambda_{ct}$ prefers to be ${\cal O}(10)$ from $B\to D^{(*)}\tau\nu$. Thus, the $\lambda_{tt}$ and $\lambda_{bb}$ must be strongly correlated to satisfy Eq.~(\ref{eq:WCconstraints}). In order to avoid large cancelation between $1/3\lambda_{tt}F_{7,8}^{(1)}$ and $\lambda_{bb}F_{7,8}^{(2)}$ in Eq.~(\ref{eq:BXsgamma:WC2HDM}) that causes fine-tuning, we prefer to take the region where $\lambda_{bb}, \lambda_{tt} \sim {\cal O}(0.1)$.

\begin{figure}[t]
  \centering
\includegraphics[width=0.6\textwidth]{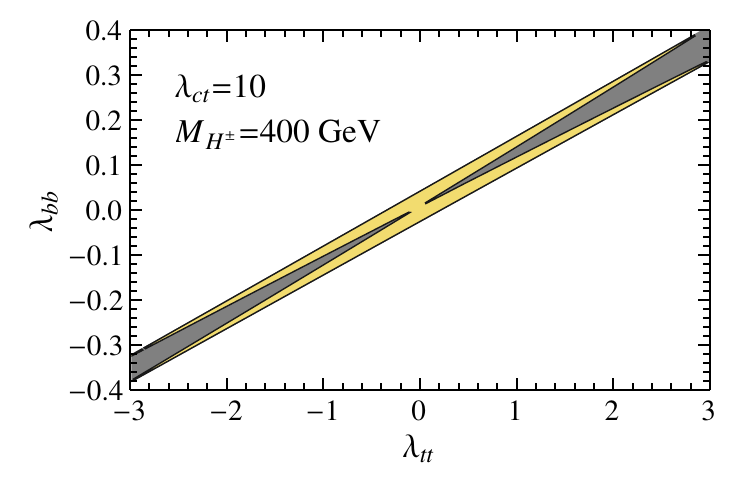}
  \caption{\baselineskip 3.0ex
  Yellow region is allowed parameter space at $95\%$\,CL from $\mathcal B(B_d \to X_s \gamma)$ with fixed $\lambda_{ct}=10$ and $M_{H^\pm}=400\GeV$ by requiring $\Delta^{-1}>10\%$. The gray region causes significant fine-tuning.}
  \label{fig:BXsgamma}
\end{figure}

    To be more specific regarding the fine-tuning argument, we refer to Ref.~\cite{Baer:2012up} and re-define fine-tuning parameter $\Delta$ for an observable as follows
 \begin{equation}
 \Delta = \frac{{\rm max}(\delta Q_i)}{Q}\,.
 \end{equation}
 Here, $Q$ denotes the difference between theory prediction and experimental data and $\delta Q_i$ represents each individual contribution of the theory to the $Q$. Therefore, small $\Delta^{-1}$ means significant fine-tuning. (For example, $\Delta=25$ correspond to $4\%$ fine-tuning.)
 The allowed parameter space in $(\lambda_{tt}, \lambda_{bb})$ plane for given $\lambda_{ct}=10$ and $M_{H^+}=400\GeV$ is shown in Fig.~\ref{fig:BXsgamma} by requiring $\Delta^{-1} > 10\%$. The gray region causes significant fine-tuning. We note that by avoiding significant fine-tuning, not only $\lambda_{tt}$ is constrained but also $\lambda_{bb}$ is highly restricted as we expected.

\subsection{$B_{d,s}-{\overline B}_{d,s}$ mixing}

The $B_{q}-{\overline B}_{q}\,(q=d,s)$ mixing occurs via box diagrams by exchanging $W$ boson or charged Higgs within 2HDM between $B_{q}$ and ${\overline B}_{q}$.
We note that the tree level diagrams can also contribute through ${\overline b}$-$s$-$(H/A)$ vertices within 2HDM type III. We first study the NP contribution from loop processes while the tree-level contribution is discussed in the next section.
The mass difference $\Delta m_{q}$ between the two mass eigenstates $B_{q}^H$ and $B_{q}^L$ is related with off-diagonal element of mixing matrix $M_{12}^q$ such that $\Delta m_{q}=2|M_{12}^q|$. Since the constraints from $B_{d}-{\overline B}_{d}$ mixing appears to be more or less weaker than those from $B_{s}-{\overline B}_{s}$ mixing, we only consider latter one in this work.
The effective Hamiltonian with $\Delta B=2$ for the $B_{s}-{\overline B}_{s}$ mixing is described by ~\cite{Buras:01a}
\begin{align}
\mathcal H^{\Delta B=2} =\frac{G_F^2}{16\pi^2}m_W^2 (V_{tb}^*V_{ts})^2\sum_i  C_i \mathcal O_i+h.c..
\end{align}
In the SM, only $\mathcal O^{\rm VLL}_1$ operator can contribute, where
\begin{align}
\mathcal O_1^{\rm VLL}=\bigl(\bar b^\alpha \gamma_\mu P_L s^\alpha\bigr) \bigl (\bar b^\beta \gamma^\mu P_L s^\beta \bigr )\,.
\end{align}
The corresponding Wilson coefficient is $C^{\rm VLL}_1(m_W) = 4S_0(x_W)$ where $x_W = m_t^2/m_W^2$. The function $S_0(x)$ can be found in Ref.~\cite{Buchalla:95}. Then the $\Delta m_{s}$ is obtained as
\begin{equation}
\Delta m_{s} = 2|\langle B_s | \mathcal H^{\Delta B=2} | {\overline B}_s \rangle | = \frac{G_F^2}{6\pi^2} |V_{tb}^*V_{ts}|^2 f_{B_s}^2 {\hat B}_{B_s} m_{B_s} \eta_b m_W^2 S_0(x_W)\,.
\end{equation}
Here, $\eta_b=0.552$ is a short-distance QCD contribution. As for the long distance non-perturbative quantity $f_{B_s} {\hat B}_{B_s}^{1/2}$, we use Lattice QCD result.

Within the 2HDM, two additional operators are generated by the box diagrams with charged Higgs boson exchanged:
\begin{align}
\mathcal O_{1}^{\rm SRR}=\bigl(\bar b^\alpha P_R s^\alpha\bigr) \bigl(\bar b^\beta P_R s^\beta\bigr),
\qquad
\mathcal O_2^{\rm SRR}=\bigl( \bar b^\alpha \sigma_{\mu\nu} P_R s^\alpha \bigr) \bigl(\bar b^\beta \sigma^{\mu\nu} P_R s^\beta \bigr).
\end{align}
Using the formulae in Ref.~\cite{Buras:01b}, the corresponding Wilson coefficients are obtained as
 \begin{align}\label{eq:Bsmixing:WC2HDM}
   C^{\rm VLL}_{1,\,HH}=&A_u^2 x_W x_{H^\pm}\biggl[\frac{x_{H^\pm}+1}{(x_{H^\pm}-1)^2}-\frac{2x_{H^\pm}\log x_{H^\pm}}{(x_{H^\pm}-1)^3}\biggr]\,,\nonumber\\
   C^{\rm VLL}_{1,\,WH}=& 2A_u x_W x_{H^\pm}\biggl[\frac{-4+x_W}{(x_{H^\pm}-1)(x_W-1)} +\frac{(x_W-4x_{H^\pm})\log x_{H^\pm}}{(x_{H^\pm}-1)^2(x_{H^\pm}-x_W)} \nonumber\\
   &\hphantom{+ 2\lambda_s x_W x_{H^\pm}\biggl(}\,+\frac{3x_W\log x_W}{(x_W-1)^2(x_{H^\pm}-x_W)}\biggr],
   \nonumber\\[0.5em]
   C^{\rm SRR}_{1,HH}=&4 A_d^2 x_{H^\pm}^2 \biggl( \frac{m_b^2}{m_W^2}\biggr) \biggl[\frac{2}{(x_{H^\pm}-1)^2} -\frac{(x_{H^\pm} +1 ) \log x_{H^\pm}}{(x_{H^\pm}-1)^3}\biggr],
  \end{align}
where $x_{H^\pm}=m_t^2/M_{H^\pm}^2$. The subscript $WH$ or $HH$ represent the exchanged particles in the box diagram.
 We note that $C_2^{\rm SRR}=0$ at the matching scale $\mu_W$. Contrary to the $B_d \to X_s \gamma$, the $A_d$ contribution in $C_1^{\rm SRR}$ has significant suppression factor $m_b^2/m_W^2$, thus its contribution is negligible. Although the operators $\mathcal O_{1}^{\rm SRR}$ and $\mathcal O_{2}^{\rm SRR}$ are generated through operator mixing during renormalization group evolution as described in detail in Refs.~\cite{Buras:90,Urban:97,Ciuchini:97a,Ciuchini:98,Buras:00,Buras:01a} at NLO QCD, the effects are minor and we do not include them.

\begin{figure}[t]
  \centering
\includegraphics[width=\textwidth]{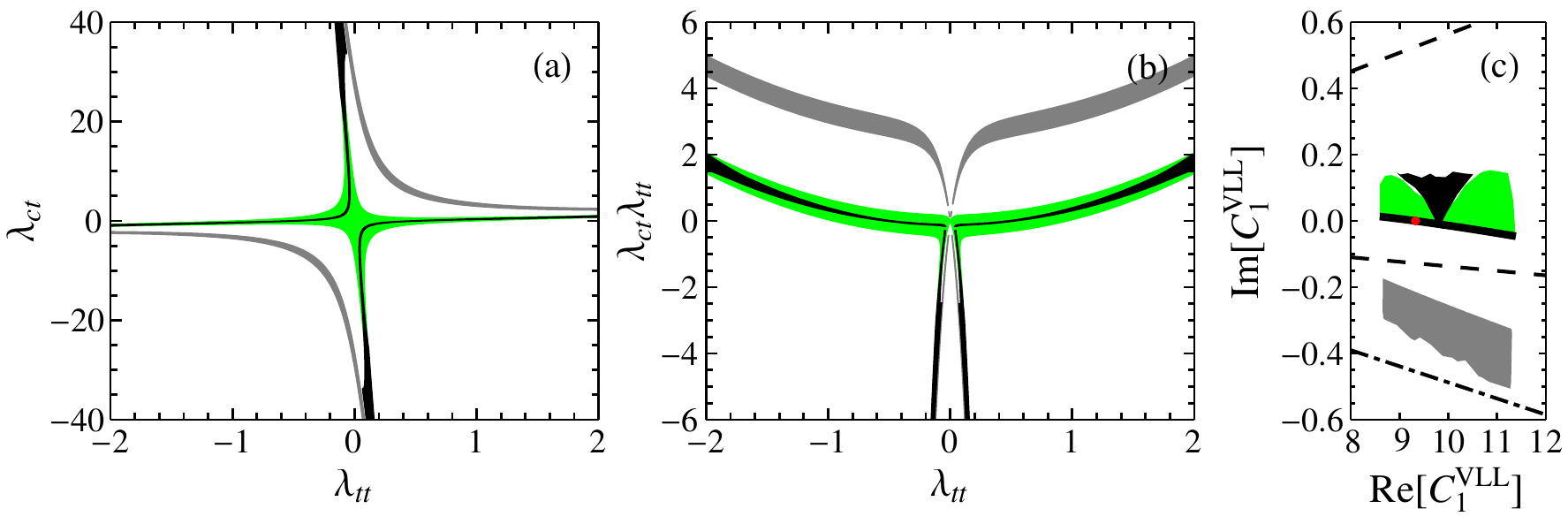}
  \caption{\baselineskip 3.0ex
  Allowed parameter space at 95\%\,{\rm CL} by $\Delta m_s$ experimental data for fixed   $M_{H^\pm}=500{\,\rm GeV}$ in (a) ($\lambda_{tt}$, $\lambda_{ct}$), (b) ($\lambda_{tt}$, $\lambda_{ct}\lambda_{tt}$) and (c) $\bigl ( {\rm Re}[C_{1}^{\rm VLL}], {\rm Im}[C_{1}^{\rm VLL} ] \bigr)$ planes. Green color (S1) corresponds to the solution without significant fine-tuning. Black color (S2) and gray color (S3) represent the parameter space with significant fine-tuning, $\Delta^{-1}<10\%$, where S3 causes large ${\rm Im} M_{12}^s$ while S2 does not. The dashed (dot dashed) line denotes $68\%$\,CL ($95\%$\,CL) bound from $\phi_s^{c \bar c s}$. The red point represents the SM prediction.}
  \label{fig:Bsmixing}
\end{figure}

Therefore, only $A_u$ is numerically relevant in $B_s - \overline B_s$ mixing. It contains  $\lambda_{tt}$ and $\lambda_{ct}$ as defined in Eq.~(\ref{eq:lambdas}) which are constrained by experimental data of $\Delta {m_s}$ given in Table~\ref{tab:exp}. The allowed region for the parameter space in $(\lambda_{tt},\lambda_{ct})$ plane as well as $(\lambda_{tt},\lambda_{ct}\lambda_{tt})$ plane are shown in Fig.~\ref{fig:Bsmixing}\,(a) and (b).
We perform more detailed study on the allowed parameter space by considering the fine-tuning argument to fit the data. As shown in Eq.~(\ref{eq:lambdas}), there are two solutions for $A_u=0$ which give the result consistent with experimental data:
\begin{eqnarray}
\lambda_{tt} &\simeq&- \frac{V_{cs}}{V_{ts}}\sqrt{\frac{m_c}{m_t}}\lambda_{ct}
\simeq (2.14-0.04\,i) \lambda_{ct} \,,\nn \\
{\rm or}~~\lambda_{tt} &\simeq& - \frac{V_{cb}^*}{V_{tb}^*}\sqrt{\frac{m_c}{m_t}}\lambda_{ct}
\simeq - 0.004 \,\lambda_{ct}\,.
\end{eqnarray}
The parameter space near these two solutions are allowed, but can cause significant fine-tuning.
We represent the allowed parameter without significant fine-tuning, or $\Delta^{-1}> 10\%$ by green color, and for $\Delta^{-1}< 10\%$ by black color.

In the region where the signs of $\lambda_{ct}$ and $\lambda_{tt}$ are same, the two 2HDM contributions $C^{\rm VLL}_{1,\,WH}$ and $C^{\rm VLL}_{1,\,HH}$ are destructive with each other.
The parameter space that brings the cancelation between the two can be another solution to fit the data, but also causes significant fine-tuning.  We represent the parameter space near the solution with significant fine-tuning,  $\Delta^{-1} < 10\%$, with gray color. For this solution space,  the real parts of the two 2HDM contributions are strongly canceled, but sizable imaginary parts still remain as can be seen in the Fig.~\ref{fig:Bsmixing}(c).
This sizable imaginary part can cause large time-dependent $CP$-asymmetry phase $\phi_s^{c \overline c s}$  in $b\to c$ decays from the relation  $\phi_s^{c \overline c s}\equiv\arg(M_{12}^s)$.
We show the bounds at $68\%$ and $95\%$ CL in Fig.~\ref{fig:Bsmixing}(c) with current average value~\cite{HFAG}
\begin{equation}
\phi_s^{c\bar c s} = -0.015\pm 0.035\,.
\end{equation}
As shown, the gray region is excluded by $\phi_s^{c\bar c s}$ at $68\%\,$CL, but survives at $95\%\,$CL. This region will be more significantly covered by future experimental data.

For later convenience, we summarize the features of each parameter regions and their color notation with the definition of S1, S2 and S3 as follows
\begin{eqnarray}
\label{eq:S123}
{\rm S1} &:&
\textrm{(green color)}~\Delta^{-1} > 10\%,
\nn\\
{\rm S2} &:&
\textrm{(black color)}~\Delta^{-1}<10\%,~A_u\simeq 0
\,, \nn\\
{\rm S3} &:&
\textrm{(gray~ color)}~\Delta^{-1} < 10\%,~{\rm Re}C^{\rm VLL}_{1,\,WH} + {\rm Re}C^{\rm VLL}_{1,\,HH}\simeq 0\,,~{\rm large~ Im} M_{12}^s
 \,.
\end{eqnarray}

\section{Combined analysis and future prospect}
\label{sec:combined}

\begin{figure}[t]
  \centering
  \includegraphics[width=\textwidth]{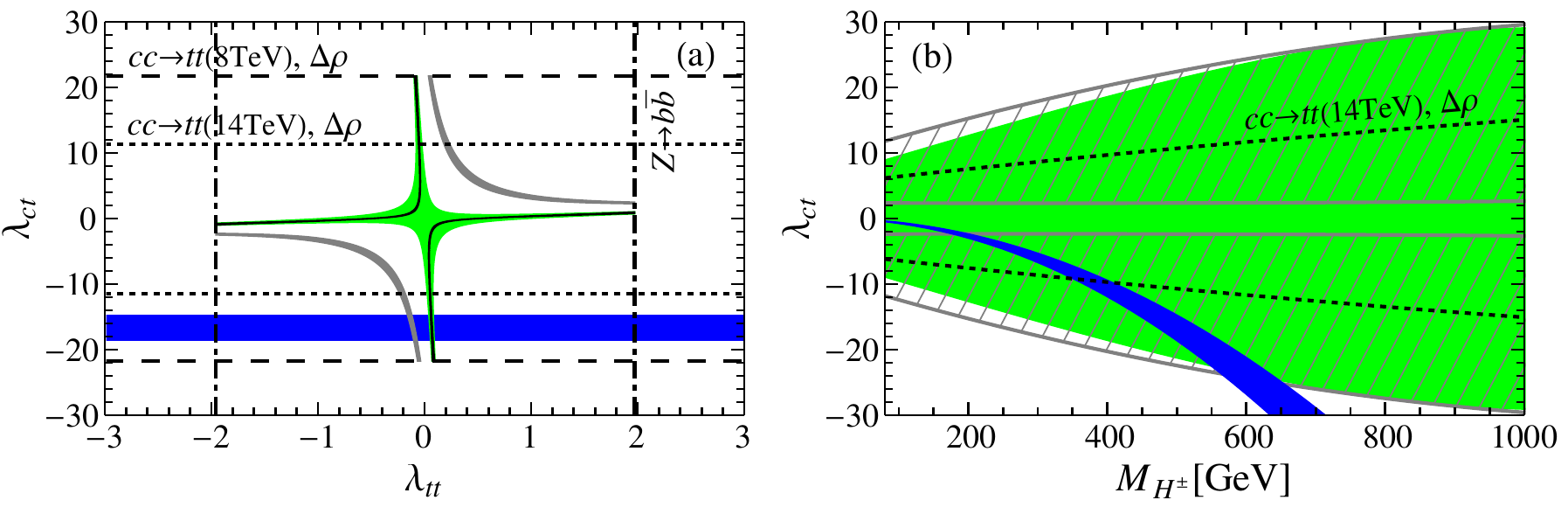}
  \caption{\baselineskip 3.0ex
  Combined constraints from $B_{s}-\overline B_{s}$ mixing, $cc \to tt$, $Z\to b \overline b$ and the oblique parameter $\Delta\rho$ on the 2HDM parameters.  The allowed regions are divided into three parts and shown in the green (S1), black (S2) and gray regions (S3). (a) Allowed parameter space in ($\lambda_{tt}$, $\lambda_{ct}$) plane for the fixed $M_{H^\pm}=500 \,{\rm GeV}$. The constraints from $cc\to tt,\Delta\rho$ and $Z\to b\overline b$ are shown in dashed and dot-dashed lines respectively. The projection for $cc\to tt, \Delta\rho$ at $14\TeV$ with $300\invfb$ data is shown by a dotted line.
    The allowed parameter space by $B\to D^{(*)}\tau\nu$ (with $\lambda_{\tau\tau}=40$) are indicated by the blue region.
  (b) Allowed parameter space in ($M_{H^\pm}$, $\lambda_{ct}$) plane.
  Note that the upper and lower bounds of black region are same with gray region so they are not shown in the plot.
}
  \label{fig:PS:All}
\end{figure}

 We first combine the constraints from $B_d \to X_s \gamma$, $B_s-\overline B_s$ mixing, and $cc \to tt$ on the couplings $\lambda_{ct}$ and $\lambda_{tt}$.  We also include the constraints from EW precision measurements, $Z\to b\overline b$ and $\Delta \rho$. We refer to Ref.~\cite{Song:2014lua} for the details of these EW precision measurements.
 We scan the parameter space as described in Eq.~(\ref{eq:paramscan}). The allowed parameter space is obtained by requiring that it accommodates all the experimental data with $95\%\,$CL. The result is shown in Fig.~\ref{fig:PS:All}(a) for $M_{H^\pm}=500{\,\rm GeV}$. As discussed in previous section we divide allowed parameter region into S1, S2 and S3 whose features are portrayed in Eq.~(\ref{eq:S123}).

  For the region S1, the requirement $\Delta^{-1}>10\%$ in $B_{s}-\overline B_{s}$ mixing gives the upper bound on  $\lambda_{ct}$ and is slightly stronger than the one from $\sigma(cc\to tt)$ combined with $\Delta\rho$.
  The upper bound on $\lambda_{tt}$ for the region S1 is given by $R_b$. On the other hand, for the regions S2 and S3, the couplings $\lambda_{ct}$ and $\lambda_{tt}$ are bounded by $\sigma(cc\to tt)$ accompanied with $\Delta\rho$ and $R_b$. Therefore, the same sign top pair production plays crucial role to constrain  $\lambda_{ct}$ regardless of fine-tuning. But if we avoid significant fine-tuning (for S1), $B_{s}-\overline B_{s}$ mixing put the significant bound.
  The projection for the exclusion limit at $14\TeV$ with $300\invfb$ is
  estimated by assuming that the statistical error is dominant (See Ref.~\cite{Jung:2013zya,Djouadi:2015jea}   for more details about the projection method). The result is outstanding. The upper bound of $\lambda_{ct}$ reach 8$\sim$15 with $300\invfb$ at $14\TeV$ as shown in Fig.~\ref{fig:PS:All}.
 We note that $B_d \to X_s \gamma$ does not put bound on $\lambda_{ct}$ nor $\lambda_{tt}$ for any parameter sets due to sizable contributions from $\lambda_{bb}$ term.

\begin{figure}[t]
  \centering
 \includegraphics[width=0.6\textwidth]{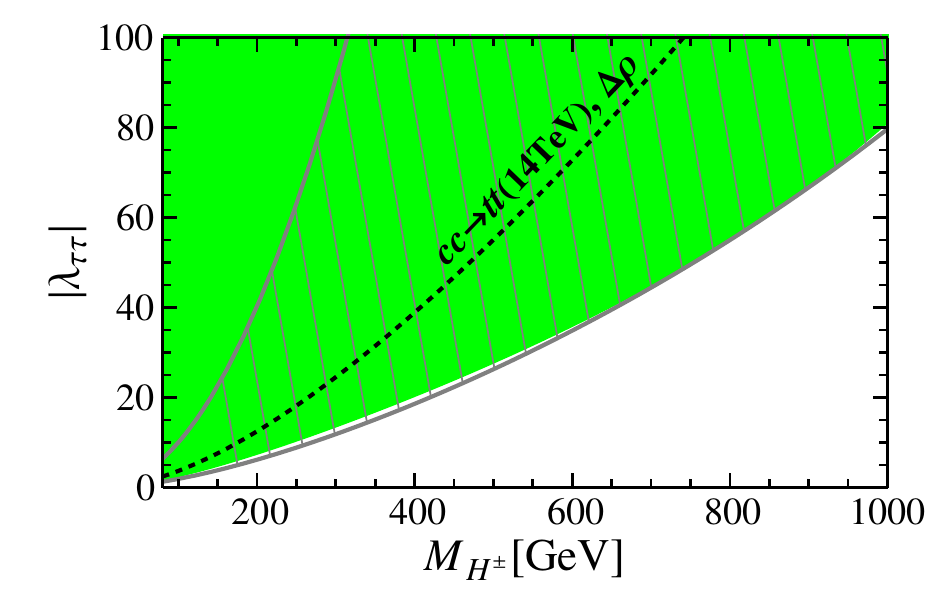}
  \caption{\baselineskip 3.0ex
  Allowed parameter space ($M_{H^\pm}$, $|\lambda_{\tau\tau}|$) by the combined constraints from loop-induced processes and $B \to D^{(*)} \tau\nu$ decays. See Eq.~(\ref{eq:S123}) for the definition of each parameter set.
  The projection for $cc\to tt, \Delta\rho$ at $14\TeV$ with $300\invfb$ data is shown by a dotted line.
  The lower bound of black region is same with gray region so is not shown in the plot.
  }
  \label{fig:PS:xiltlt}
\end{figure}

We turn to the $B\to D^{(*)}\tau\nu$ decays. With fixed $\lambda_{\tau\tau}$, $B\to D^{(*)}\tau\nu$ decays also put bounds on $M_{H^\pm}$ and $\lambda_{ct}$. By taking $\lambda_{\tau\tau}=40$, the allowed parameter space is shown in blue-colored region in Fig.~\ref{fig:PS:All} (with $M_{H^\pm}=500\,\rm GeV$). As shown in Fig.~\ref{fig:PS:All}(b), $|\lambda_{ct}|$ has different upper limits for each parameter set depending on $M_{H^\pm}$. They lead to lower limits on $|\lambda_{\tau\tau}|$ as can be seen in Eq.~(\ref{eq:treebound}) and Fig.~\ref{fig:treelevel}. The allowed parameter spaces in ($M_{H^\pm}$, $|\lambda_{\tau\tau}|$) plane are presented in Fig.~\ref{fig:PS:xiltlt}. For fixed $M_{H^\pm}$, the lower bounds for S2 and S3 are same and slightly different from S1.
It should be noted that these lower bounds become stronger as $M_{H^\pm}$ increases. Conversely, the $M_{H^\pm}$ is upper bounded when $\lambda_{\tau\tau}$ is fixed.
In the case of relatively  heavy charged Higgs, the lower bound on $\lambda_{\tau\tau}$ is very strong. With the constrains of $cc\to tt$ at $14\TeV$ with $300\invfb$ data, the lower bound on $\lambda_{\tau\tau}$ would become twice of current bound as shown in Fig.~\ref{fig:PS:xiltlt}.
  For $M_{H^\pm}>500 {\,\rm GeV}$, the coupling $\lambda_{\tau\tau}$ should be greater than 30, which can significantly enhance $H/A \to \tau\tau$ decays.
Therefore, this can be constrained by heavy Higgs search with $\tau\tau$ final states at the LHC.
However the signal strength of $gg\to H/A \to \tau\tau$ process strongly depends on heavy Higgses masses and is effectively proportional to $\lambda_{tt}^2$. Since there are much parameter space near $\lambda_{tt}\sim 0 $ in the set S1 (green region) as shown in Fig.~\ref{fig:PS:All} that may avoid the constrains from $gg\to H/A \to \tau\tau$, the constraints would be restricted. Perhaps, some part of parameter space, especially small $\lambda_{ct}$ and large $\lambda_{tt}, \lambda_{\tau\tau}$ region will be excluded. On top of that, for such very large $\tau$ Yukawa coupling, the perturbativity would be threatened.

\begin{figure}[t]
  \centering
\hspace{-0.33cm} \includegraphics[width=0.6\textwidth]{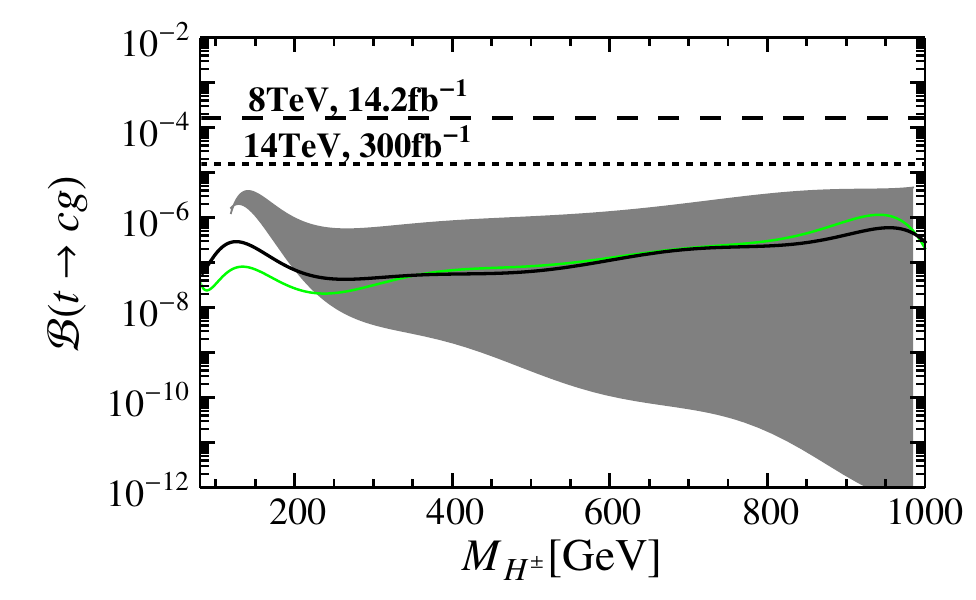}
  \caption{\baselineskip 3.0ex
  2HDM prediction on $\mathcal B(t \to c g)$ as a function of $M_{H^\pm}$. The green and black lines denote the upper bounds from the solutions of the combined constraints S1 and S2, respectively. The gray region corresponds to the solution S3. The dashed line denotes the current upper bound at LHC, while the dotted line is for the future sensitivity at $14\TeV$ with $300\,{\rm fb}^{-1}$ data.
  }
  \label{fig:BR:tcg:2HDM}
\end{figure}

 We now discuss about the constraints from $t \to cg$. With the above allowed regions S1, S2 and S3, we make theoretical predictions for $\Br(t \to cg)$.
Since the combined constraints put upper bounds on both $\lambda_{ct}$ and $\lambda_{tt}$, Therefore, $\lambda_{ct}\lambda_{tt}$ is upper bounded in all three parameter sets S1, S2 and S3. Note that the set S3 represents also the lower bounds for both $\lambda_{ct}$ and $\lambda_{tt}$ that comes from $Z\to b \overline b$ and $cc\to tt$ as shown Fig.~6(a). The upper bound of  $\mathcal B(t \to cg)$  for S1, S2 and the allowed region for S3 are presented as a function of $M_{H^\pm}$ in Fig.~\ref{fig:BR:tcg:2HDM}.

The current LHC upper limit is much larger than these theory predictions. Thus, it does not give any constraints. The projection for the upper limit at $14\TeV$ with $300\invfb$ data is also drawn in Fig.~\ref{fig:BR:tcg:2HDM} in dotted line. As shown, it would be hopeless to see or constrain the top quark FCNC couplings from the $ t\to cg$ measurement.

\begin{figure}[t]
  \centering
 \hspace{-0.33cm} \includegraphics[width=0.6\textwidth]{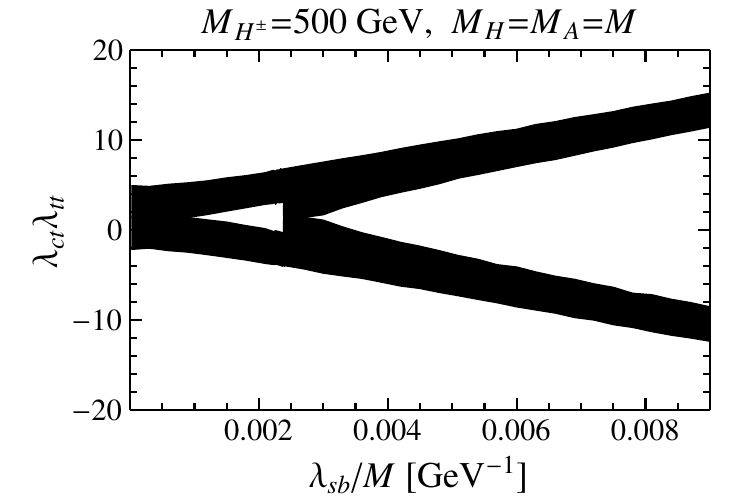}
  \caption{\baselineskip 3.0ex
  Allowed parameter space in $(\lambda_{sb}/M,\lambda_{ct}\lambda_{tt})$ by the combined constraints from $B_s-\overline B_s$ mixing, $cc\to tt$, $Z \to b\overline b$ and the oblique parameter $\Delta\rho$. $M_H=M_A=M$ and $M_{H^\pm}=500\GeV$ are taken. For $B_s-\overline B_s$ mixing, both the tree-level and loop-level contributions are included. }
  \label{fig:lamsb}
\end{figure}

So far, we have neglected the tree-level contribution to $B_s-\overline B_s$ mixing through the down-type FCNC couplings $\overline b$-$s$-$(H/A)$ with the Yukawa coupling $\xi_{bs}$. Even though $\xi_{bs}$ is severely suppressed in Cheng-Sher ansatz such as $\xi_{bs}/\lambda_{bs}=3.6\times 10^{-3}$, the tree level contribution with ${\cal O}(1)~\lambda_{sb}$ has no CKM suppression, and is comparable to the loop contribution. By including the tree level contribution, the allowed parameter space in $(\lambda_{tt},\lambda_{ct})$ plane is significantly extended since the large NP contribution from the loop processes can be canceled by the tree level contribution. Therefore, including the tree level contribution in $B_s-\overline B_s$ mixing always weakens the constraints on $\lambda_{tt},\lambda_{ct}$.
To understand the effect of $\lambda_{sb}$ quantitatively, we show a plot in Fig.~\ref{fig:lamsb} for allowed region of $\lambda_{ct}\lambda_{tt}$ with respect to the fixed $\lambda_{sb}$ value by imposing the constraints from $cc\to tt$, $Z\to b \overline b$ and $\Delta \rho$. We see that for $\lambda_{sb} > 0.003 M_{H,A}$ large $\lambda_{ct}\lambda_{tt}$ is required to cancel the large tree-level contribution. In fact, for $\lambda_{sb} \simeq 0.003 M_{H,A}$, the magnitude of tree-level contributions is already comparable to the magnitude of the SM contributions. For $\lambda_{sb} < 0.003 M_{H,A}$ , the bound on $\lambda_{ct}\lambda_{tt}$ is not much changed from the one given in previous section.

\section{Conclusion}
The general 2HDM as an extension to the SM is a potential NP candidate. To avoid severe constraints from down-type
quark FCNC, we adopt Cheng-Sher ansatz. This NP scenario permits
presumably large top quark FCNC coupling $\lambda_{ct}$, which is
the main target to be explored in this work with collider
phenomenology as well as flavor constraints and EW precision
measurements.
 To this end, we consider anomalous single top production
 which can limit $\Br(t\to cg)$ and the same sign top pair production via $cc\to tt$ at the LHC in association
 with not only flavor tree-level processes,  $B \to D^{(*)}\tau\nu$, $B\to \tau\nu$ but also flavor loop-level processes, $B_d \to X_s \gamma$, $B_{s}-\overline B_{s}$ mixing.

 We find that among them the $B \to D^{(*)}\tau\nu$, $B_{s}-\overline B_{s}$ mixing and $cc\to tt$ play important role to constrain $\lambda_{ct}$. Especially, still large value of $\lambda_{ct}$ is preferred by average value of $R(D^{(*)})$ measurement with the new data for $B \to D^{(*)}\tau\nu$ from Belle and LHCb. To bring solid understanding of the result, we separate the allowed parameter space into three sets, S1, S2 and S3, regarding the fine-tuning to fit the data and the features reflected in the observables of $B_{s}-\overline B_{s}$ mixing. S1 does not suffer from the fine-tuning while S2 and S3 cause significant fine-tuning to fit the data. More specifically, S3 shows large imaginary part of $M_{12}^s$ while S1 and S2 do not.

  For the allowed parameter sets S1, S2 and S3, $\lambda_{ct}$ is severely upper-bounded by either
  $cc\to tt$ or $B_{s}-\overline B_{s}$ mixing.
  Therefore, to fit the $R(D^{(*)})$ values, the Yukawa coupling $\lambda_{\tau\tau}$ is lower bounded for given charged Higgs mass $M_{H^\pm}$ and conversely $M_{H^\pm}$ is upper bounded for fixed $\lambda_{\tau\tau}$. The large $\lambda_{\tau\tau}$ will be constrained by $gg\to H/A \to \tau\tau$, however it strongly depends on neutral Higgses masses and $\lambda_{tt}$. The extended study with heavy Higgs search data at the LHC can be a future work.
  Since $\lambda_{ct}\lambda_{tt}$ is small for all the parameter sets and the theory prediction is loop-suppressed, the upper limits for $\Br(t\to cg)$
  do not provide constraints on the remaining parameter space with current experimental data nor in future LHC experiment.
   On the other hand, large $\lambda_{ct}$ is  mostly constrained by $cc\to tt$ process regardless of fine-tuning.
  $cc\to tt$ would play more important role to probe top quark FCNC at the LHC $14\TeV$ Run.

\begin{acknowledgements}
CSK and XY are  supported by the NRF grant funded by the Korean
government of the MEST (No. 2011-0017430) and (No. 2011-0020333).
YWY is supported in part by NRF-2013R1A1A2061331 and in part by
NRF-2012R1A2A1A01006053. YWY thank Sunghoon Jung for useful
discussions. We thank KIAS Center for Advanced Computation for
providing computing resources.
\end{acknowledgements}

\begin{appendix}

\section{Form factors in $t \to cg$}
\label{sec:FF}

In general 2HDM, the form factors for $tcg$ vertex was first
calculated in Refs.~\cite{Luke:93,Atwood:96}. Here, we recalculate
these form factors and write them in terms of scalar one-loop
functions. Each form factor in Eq.~(\ref{eq:formfactor}) is
summation of four different contributions from the penguin
diagrams with $A$, $H$ and $H^{\pm}$ exchanges, e.g. ${\cal
A}={\cal A}_A+{\cal A}_H+{\cal A}_{H^\pm}$. They are calculated as
\begin{align}
  {\cal A}_A &=-g_s \xi_A^A f_1^A, &
  {\cal A}_H&=g_s \xi_H^V f_1^H, &
  {\cal A}_{H^\pm}&=g_s |V_{tb}|^2 \xi_{H^\pm} f_1^{H^\pm},
  \nonumber\\
  {\cal B}_A &= g_s \xi_A^V f_1^A,&
  {\cal B}_H &= -g_s \xi_H^A f_1^H,&
  {\cal B}_{H^\pm}&=g_s |V_{tb}|^2 \xi_{H^\pm} f_1^{H^\pm},
  \nonumber\\
  {\cal C}_A &= -g_s \xi_A^A f_2^A,&
  {\cal C}_H&= g_s   \xi_H^V f_2^H,&
  {\cal C}_{H^\pm} &=  g_s |V_{tb}|^2  \xi_{H^\pm} f_2^{H^\pm},
  \nonumber\\
  {\cal D}_A &= -g_s \xi_A^V f_2^A,&
  {\cal D}_H&= g_s \xi_H^A f_2^H,&
  {\cal D}_{H^\pm} &= - g_s |V_{tb}|^2  \xi_{H^\pm}  f_2^{H^\pm}\,.
\end{align}
To compare with Refs.~\cite{Luke:93,Atwood:96}, we neglect the
small term $V_{cb} \xi_{ct}$ in $\overline t b H^+$ vertex of
Eq.~(\ref{eq:YukawaL2}) and show the result in general with
complex Yukawa couplings
\begin{align}
\label{eq:xidef}
  \xi_H^V&=\frac{1}{4}\xi_{tt}(\xi_{ct}+\xi_{tc}^*),&
  \xi_A^A&=\frac{1}{4}\xi_{tt}(\xi_{ct}-\xi_{tc}^*),&
  \xi_{H^\pm}=\frac{1}{4}\xi_{ct}\xi_{tt},
  \nonumber\\
  \xi_H^A&=\frac{1}{4}\xi_{tt}(\xi_{ct}-\xi_{tc}^*),&
  \xi_A^V&=\frac{1}{4}\xi_{tt}(\xi_{ct}+\xi_{tc}^*).&
\end{align}
The loop functions are defined as
\begin{flalign}
\qquad\qquad
f_1^A&=  q^2( C_0^A -     2C_{11}^A  -   C_{12}^A +   C_2^A),&
f_2^{A}&=m_t^2( C_0^A      -  C_{12}^A  +  C_2^A ),&
\nonumber\\
f_1^H&=q^2(C_0^H + 2C_{11}^H +C_{12}^H+ C_2^H + 4 C_1^H  ),&
f_2^H&=m_t^2(C_0^H +    C_{12}^H  + C_2^H ),&
\nonumber\\
f_1^{H^\pm}&=q^2( 4  C_1^{H^\pm}      + 4 C_{11}^{H^\pm}  + 2 C_{12}^{H^\pm}),&
f_2^{H^\pm}&=m_t^2 (2C_{12}^{H^\pm} ).  &
\end{flalign}
The scalar one-loop functions are abbreviated as
\begin{align}
  C_{ij}^{H,A}=C_{ij}( q^2, m_t^2, 0, m_t^2, m_t^2, m_{H,A}^2),
  \quad\quad
  C_{ij}^{H^\pm}&=C_{ij}(q^2, m_t^2, 0, 0, 0, M_{H^\pm}^2),
\end{align}
which are defined in Refs.~\cite{Passarino:78,'tHooft:78,Hahn:98}
and can be numerically evaluated by the \texttt{LoopTools}
package~\cite{Hahn:98}. In the penguin diagrams with charged Higgs
$H^\pm$, we omit the terms proportional to $\xi_{bb}$ as in
Refs.~\cite{Luke:93,Atwood:96}, since these terms are suppressed
by $m_b/v$. In addition, we have analytically checked that the
form factors presented in this paper are in agreement with those
obtained in Ref.~\cite{Atwood:96} except one minor discrepancy:
for the parameter $\beta^{H,A}$ defined in Ref.~\cite{Atwood:96},
we obtained $\beta^{H,A} = x^2 m_t^2 + (1-x)M_{H,A}^2$. But this does not come into play in our numerical analysis.

\end{appendix}

\end{document}